\documentclass{article}
\usepackage{spconf,amsmath,amssymb,booktabs,graphicx,multirow,array}
\makeatletter
\def\name#1{\gdef\@name{{\em #1}}}
\makeatother

\usepackage{url}
\usepackage{hyperref}
\usepackage{xcolor}
\usepackage{colortbl}
\usepackage{makecell}

\usepackage[table]{xcolor}
\definecolor{stagefinal}{gray}{0.94}

\usepackage{etoolbox}
\makeatletter
\patchcmd{\thebibliography}
  {\list}
  {\fontsize{8.5pt}{9.5pt}\selectfont
   \list}
  {}{}
\apptocmd{\thebibliography}{%
  \setlength{\itemsep}{3pt}%
  \setlength{\parsep}{0pt}%
  \setlength{\parskip}{0pt}%
}{}{}
\makeatother

\title{UNITE-AUDIO: Joint Learning of Continuous Tokenization and \\ Latent Flow Matching for Text-to-Audio Generation}

\name{\shortstack[c]{
Runwu Shi$^{1}$, Kai Li$^{2}$, 
Yujin Wang$^{3}$, 
Dong Yang$^{4}$, 
Jiahui Li$^{1}$, 
Jiang Wang$^{1}$, 
Chang Zeng$^{4}$, 
\\[-0.2em]
Benjamin Yen$^{1}$, 
Ashizawa Takeshi$^{1}$,
Chunxiang Jin$^{5}$, 
Kazuhiro Nakadai$^{1}$
}}
\address{
$^{1}$Institute of Science Tokyo,
$^{2}$Tsinghua University,
$^{3}$Wuhan University,
$^{4}$The University of Tokyo,
$^{5}$Ant Group
}
\begin{document}
\maketitle

\begin{abstract}
Text-to-audio (TTA) generation aims to synthesize realistic audio that
faithfully reflects natural-language descriptions. Most TTA systems adopt a
two-stage latent paradigm: an audio tokenizer is optimized for reconstruction
and then frozen, after which a generative model is trained in the resulting
latent space. However, reconstruction-oriented representations may be
suboptimal for generation, motivating joint representation and generative
learning. To this end, we introduce \textbf{Unite-Audio}, to our knowledge, is the \textbf{first} to jointly learn continuous audio
representations and latent flow matching for TTA.
By coupling reconstruction with self-supervised generative prediction,
Unite-Audio allows the generative objective to directly shape the latent space
rather than treating it as a fixed intermediate representation. We further
employ Flow-GRPO post-training to improve text-conditioned generation.
Experiments show competitive TTA performance with a compact latent flow model,
while ablation studies confirm the benefit of jointly learning the audio
representation and generative model. Audio samples and checkpoints are available at
\url{https://runwushi.github.io/Unite-Audio/}.
\end{abstract}


\section{Introduction}
\vspace{-0.5em}

\label{sec:introduction}
Recent text-to-audio (TTA) systems have made substantial progress in generating
diverse acoustic scenes and sound events from open-ended textual descriptions.
Modern TTA systems predominantly adopt latent generative modeling, where
high-dimensional audio signals are compressed into compact representations and
a conditional diffusion or flow model is trained to generate within this latent space
\cite{liu2023audioldm,liu2023audioldm2,hung2024tangoflux}. 

However, in this two-stage paradigm, the audio tokenizer is typically optimized for reconstruction independently of the downstream generative objective. As a result, the learned latent space, while effective for faithful decoding, may not necessarily be the most suitable representation for generative modeling. To improve the generative properties of reconstruction-oriented latents, recent speech generation systems have incorporated semantic supervision from pretrained language or automatic speech recognition (ASR) models into tokenizer training. For example, Qwen-Audio-3.0-Gen leverages a pretrained language model \cite{dai2026qwenaudio}, while Ming-UniAudio introduces semantic knowledge from a Whisper encoder to shape their continuous audio representations \cite{yan2025minguniaudio}. 

\begin{figure}[!tb]
  \centering
  \centerline{\includegraphics[width=8.5cm]{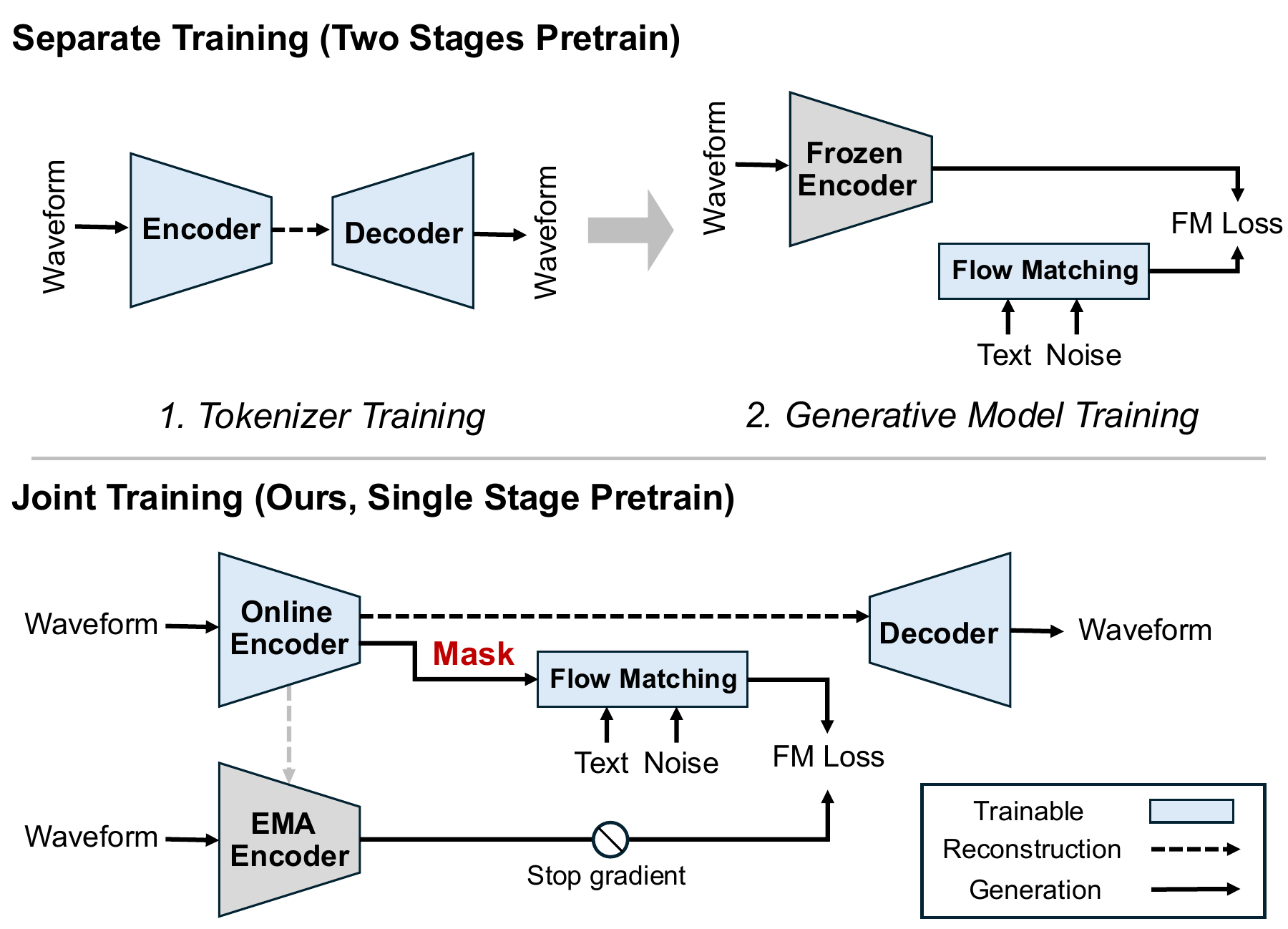}}
%  \vspace{2.0cm}
  % \centerline{(a) Result 1}\medskip
\caption{Comparison between conventional separate tokenizer–generator training and our joint single-stage training.}
\label{fig_01}
\vspace{-1em} 
\end{figure}

Beyond aligning tokenizers with external semantic models, recent studies in visual generation have explored shaping the latent space directly through self-supervised and generative objectives. MAETok introduces masked modeling into tokenizer training, showing that more structured latent representations can facilitate downstream generative modeling while maintaining reconstruction fidelity \cite{chen2025maetok}. More recent approaches further couple representation learning with latent generative modeling. Unified Latents regularizes learned representations with a diffusion prior \cite{heek2026unified}, while UNITE jointly optimizes tokenization and latent denoising in a single-stage framework, allowing the generative objective to directly participate in shaping the latent space \cite{duggal2026unite}. Meanwhile, recent advances in visual generation have shown that self-supervised representations can provide strong priors for generative modeling. Methods such as REPA \cite{yu2025repa}, RAE \cite{zheng2025rae}, and ReDi \cite{kouzelis2025redi} exploit pretrained representations or jointly model semantic features to improve generative performance. These studies suggest that generation can benefit not only from better optimization of the generative model, but also from a latent space endowed with stronger representation structure.

Building on these two complementary directions, we introduce
\textbf{Unite-Audio}, which, to the best of our knowledge, is the first TTA
framework to jointly optimize continuous audio tokenization and latent
generative modeling. Unite-Audio further incorporates self-supervised
prediction into latent representation learning and learns its latent space
entirely from scratch, without relying on external pretrained models for
representation alignment. Specifically, the online encoder observes a partial
waveform and produces latent representations that are noised and fed to the
latent flow model, while an exponential-moving-average (EMA) encoder observes
the full waveform and provides the full latent target. By predicting the full
EMA representation, the flow-matching objective can propagate through the
online representations and directly shape the encoder, while reconstruction
preserves the information required for faithful decoding. We additionally
interleave full-generation training from Gaussian noise to match the inference
setting. An overview of the proposed framework is illustrated in
Figure~\ref{fig_01}. Experimental results and ablation studies demonstrate the
effectiveness of these designs for learning a generation-friendly latent space,
with Unite-Audio achieving competitive TTA performance despite its compact
model size.

\vspace{-0.5em}
\section{Unite-Audio Framework}
\vspace{-0.5em}
\subsection{Overview}
\vspace{-0.5em}

\begin{figure}[!tb]
  \centering
  \centerline{\includegraphics[width=8.5cm]{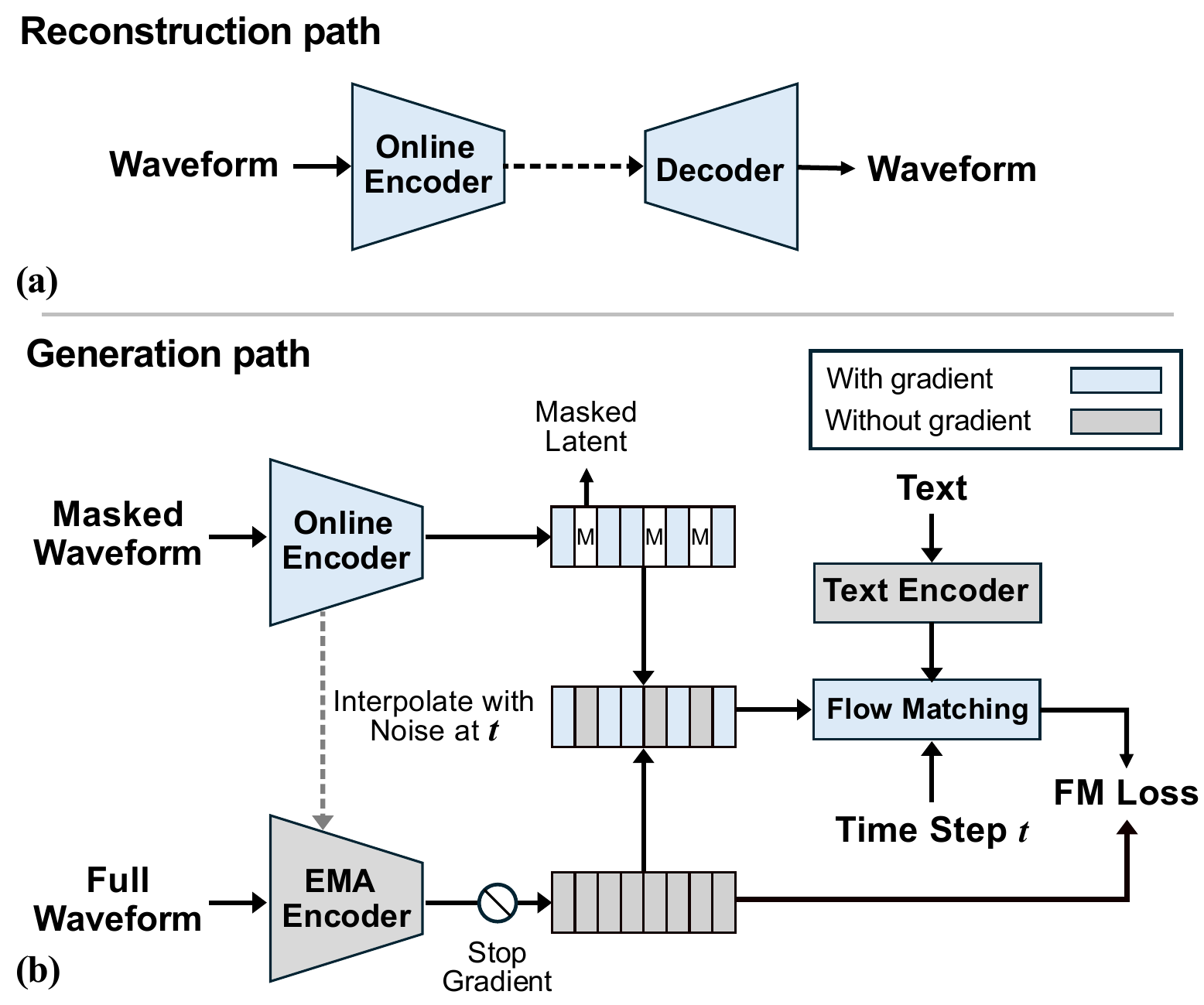}}
\caption{Overview of the proposed framework: (a) the reconstruction path and
(b) the generation path. Blue denotes regions with gradient propagation, while
gray denotes regions without gradient propagation.}
\label{fig_02}
\vspace{-1em}
\end{figure}

An overview of the training framework of Unite-Audio is illustrated in Figure~\ref{fig_02}. Let
$x\in\mathbb{R}^{T}$ denote an audio waveform and $c$ its caption. Unite-Audio
consists of an online encoder $E_{\theta}$, a decoder $D_{\phi}$, an EMA encoder $\bar{E}_{\theta}$, and a
text-conditioned latent flow model $F_{\psi}$. During training, a fraction
$p_{\mathrm{rec}}$ of each batch follows the reconstruction path, while the
remaining $1-p_{\mathrm{rec}}$ follows the generation path; both paths
propagate gradients to $E_{\theta}$ and jointly shape the latent space.

For reconstruction, the full waveform is encoded as
$z_{\mathrm{rec}}=E_{\theta}(x)$ and decoded by $D_{\phi}$. For generation,
the online encoder observes only partial waveform context, while the EMA encoder
observes the full waveform and provides the stop-gradient target
$z^{\star}=\operatorname{sg}(\bar{E}_{\theta}(x))$. The partial online
representations are incorporated into the flow-model input while retaining
their gradient connection to $E_{\theta}$, allowing the generative objective
to directly participate in representation learning.

\vspace{-0.5em}
\subsection{Partial-Context Flow Matching}
\vspace{-0.5em}

For the generation path, we divide each waveform into non-overlapping
2-second temporal buckets and independently construct partial context within
each bucket. Specifically, the context ratio $r_{\mathrm{ctx}}$ of each bucket
is independently sampled between $0$ and $30\%$, where the context refers to
the retained (unmasked) waveform regions. Given the resulting contextual
waveform $x_{\mathrm{ctx}}$, the online encoder produces
$z^{\mathrm{ctx}}=E_{\theta}(x_{\mathrm{ctx}})$, while the EMA encoder observes
the full waveform and provides the full-view target $z^{\star}$.

To reduce the discrepancy between partial-context training and generation from
noise, we noise the contextual representation at the same flow time $t$ as the
EMA trajectory. Given $\epsilon\sim\mathcal{N}(0,I)$, we define
$z_t^{\star}=(1-t)\epsilon+t z^{\star}$ and
$z_t^{\mathrm{ctx}}=(1-t)\epsilon+t z^{\mathrm{ctx}}$.
Let $\mathcal{C}$ denote the contextual positions. The complete flow input is
constructed as
\begin{equation}
\tilde z_{t,i} =
\begin{cases}
z_{t,i}^{\mathrm{ctx}}, & i\in\mathcal{C},\\
z_{t,i}^{\star}, & i\notin\mathcal{C}.
\end{cases}
\label{eq:context_routing}
\end{equation}
Thus, $F_{\psi}$ always receives a complete noisy latent sequence, while only
the contextual representations retain a gradient path to $E_{\theta}$.
Unlike conventional masked self-supervised learning, using clean online context
together with noisy flow states leads to a mismatch and degrades generation
performance. We therefore noise the online context to the same flow time $t$.

We adopt an $x$-prediction flow-matching objective, using the clean full-view
EMA representation as the target:
\begin{equation}
\mathcal{L}_{\mathrm{FM}}
=
\mathbb{E}_{t,\epsilon}
\left[
\frac{1}{|\bar{\mathcal C}|}
\sum_{i\notin\mathcal C}
\left\|
F_{\psi}(\tilde z_t,c,t)_i-z_i^{\star}
\right\|_2^2
\right],
\label{eq:fm_loss}
\end{equation}
where the loss is evaluated over the target regions complementary to the
online context. Following UNITE~\cite{duggal2026unite}, we use a shifted
flow-time schedule that emphasizes high-noise states. Through the contextual
positions in $z_t^\mathrm{ctx}$, $\mathcal{L}_{\mathrm{FM}}$ can propagate to the
online encoder.

\vspace{-0.5em}
\subsection{Training Objectives and Schedule}
\vspace{-0.5em}

The reconstruction objective follows recent audio autoencoder training recipes
\cite{parker2026same,evans2026stableaudio3}. We combine global spectral
reconstruction, local multi-resolution spectral reconstruction, and adversarial
learning. The global STFT loss $\mathcal{L}_{\mathrm{GSTFT}}$ is computed over
the full valid waveform to preserve global spectral structure. The phase-aware
MRSTFT loss $\mathcal{L}_{\mathrm{MRSTFT}}$ is computed on random 2-second
segments using seven dyadic FFT sizes from 32 to 2048. We further use a GAN
loss $\mathcal{L}_{\mathrm{GAN}}$ with multi-view STFT/PQMF discriminators and
a feature-matching loss $\mathcal{L}_{\mathrm{feat}}$ to improve perceptual
realism and stabilize adversarial training \cite{evans2026stableaudio3}. The reconstruction objective is
defined as
\begin{equation}
\mathcal{L}_{\mathrm{rec}}
=
\lambda_{\mathrm{G}}\mathcal{L}_{\mathrm{GSTFT}}
+
\lambda_{\mathrm{M}}\mathcal{L}_{\mathrm{MRSTFT}}
+
\lambda_{\mathrm{GAN}}\mathcal{L}_{\mathrm{GAN}}
+
\lambda_{\mathrm{F}}\mathcal{L}_{\mathrm{feat}} .
\label{eq:rec_loss}
\end{equation}
Here, $\lambda_{\mathrm{G}}$, $\lambda_{\mathrm{M}}$,
$\lambda_{\mathrm{GAN}}$, and $\lambda_{\mathrm{F}}$ control the corresponding
loss contributions. The discriminator is optimized separately using
$\mathcal{L}_{\mathrm{D}}$.

For latent generation, we use the $x$-prediction flow-matching loss
$\mathcal{L}_{\mathrm{FM}}$ defined in the previous section. With
$p_{\mathrm{rec}}$ denoting the probability of reconstruction routing, the
joint objective is
\begin{equation}
\mathcal{L}_{\mathrm{total}}
=
p_{\mathrm{rec}}\mathcal{L}_{\mathrm{rec}}
+
(1-p_{\mathrm{rec}})\mathcal{L}_{\mathrm{FM}} .
\label{eq:joint_loss}
\end{equation}
The reconstruction objective maintains acoustic fidelity and decodability,
while flow matching provides predictive and generative supervision to the latent representation.

\textbf{Stage I: Joint latent training.}
We jointly optimize $E_{\theta}$, $D_{\phi}$, and $F_{\psi}$ using
$\mathcal{L}_{\mathrm{total}}$ from scratch. Partial-context flow matching allows
$\mathcal{L}_{\mathrm{FM}}$ to propagate to the online encoder, while
full-generation examples expose $F_{\psi}$ to generation directly from noise.
The EMA encoder $\bar{E}_{\theta}$ receives no gradients and is updated only
through exponential moving averaging.

\textbf{Stage II: Full-generation training.}
We freeze $E_{\theta}$, $D_{\phi}$, and $\bar{E}_{\theta}$ and optimize only
$F_{\psi}$ using $\mathcal{L}_{\mathrm{FM}}$. Reconstruction and
partial-context training are removed, specializing the flow model to
text-conditioned generation from Gaussian noise while keeping the latent
representation fixed.

\textbf{Stage III: Flow-GRPO post-training.}
Finally, we further optimize $F_{\psi}$ using Flow-GRPO
\cite{liu2025flowgrpo}, while keeping the encoder and decoder fixed. For each
caption, multiple generations are sampled to form a group. The composite reward
is defined as
$R_i=\lambda_{\mathrm{CLAP}}z_g(r_i^{\mathrm{CLAP}})
+\lambda_{\mathrm{PaSST}}z_g(-r_i^{\mathrm{PaSST}})$,
and the group-relative advantage is $A_i=z_g(R_i)$, where $z_g(\cdot)$ denotes
within-group normalization. Here, $r_i^{\mathrm{CLAP}}$ is the CLAP similarity,
while $r_i^{\mathrm{PaSST}}$ is the PaSST KL divergence between the generated
audio and its paired reference audio from the training set. The policy is
optimized using the clipped Flow-GRPO objective:
\begin{equation}
\begin{aligned}
\mathcal{L}_{\mathrm{GRPO}}
=&~\mathbb{E}_{i,t}\!\left[
-\min\left(
\rho_{i,t}A_i,\,
\operatorname{clip}(\rho_{i,t},1-\epsilon,1+\epsilon)A_i
\right)\right] \\
&+\beta D_{\mathrm{KL}}
\left(\pi_{\psi}\Vert\pi_{\mathrm{ref}}\right),
\end{aligned}
\end{equation}
where $\rho_{i,t}$ is the transition likelihood ratio between the current and
rollout policies, and $\pi_{\mathrm{ref}}$ is the frozen Stage-II flow model.

\vspace{-0.5em}
\section{Experiments}
\vspace{-0.5em}
\subsection{Dataset}
\vspace{-0.5em}

Unite-Audio uses three audio-caption datasets across the training stages.
Stage I uses 1,729,107 AudioSet examples
\cite{gemmeke2017audioset} and 246,324 WavCaps examples
\cite{mei2023wavcaps}, with all AudioCaps test-set entries explicitly removed
from the AudioSet-derived data to prevent evaluation leakage. Stage II and
Stage III use the AudioCaps training set \cite{kim2019audiocaps}. We retain
WavCaps samples shorter than 20 seconds and resample all audio to 16 kHz with a
maximum duration of 20 seconds. Before encoding, waveforms are scaled by a fixed
factor of 8, and the resulting latent representations are RMS-normalized to a
fixed scale of 1.0, such that the flow model operates on a unit-RMS latent
space.

\vspace{-0.5em}
\subsection{Model Architecture}
\vspace{-0.5em}

Unite-Audio consists of a continuous audio tokenizer and a text-conditioned
latent flow model. The tokenizer operates on 160-sample waveform patches and
produces 64-dimensional latent representations at 25 Hz. The encoder and decoder each contain 12 same-resolution Transformer blocks, followed by two downsampling and upsampling blocks, respectively. The latent flow model follows the multimodal diffusion transformer (MMDiT)
architecture \cite{esser2024scaling}, where audio and text representations interact through
joint attention blocks. The flow model uses a hidden dimension of 512 and 18
Transformer layers, including 11 fused audio--text MMDiT blocks. Text
conditioning is provided by a frozen FLAN-T5-Large encoder
\cite{chung2022scaling}. The trainable latent flow model contains 117.2M
parameters, excluding the tokenizer and frozen text encoder.

\begin{table*}[t]
  \centering
  \caption{Comparison under the TangoFlux AudioCaps-886 protocol.
Params. counts only the parameters of the latent generative model, excluding the audio encoder--decoder and text encoder.
$^{\ddagger}$ Results are directly taken from the TangoFlux benchmark
\cite{hung2024tangoflux}; $^{\dagger}$ denotes our evaluation of official
checkpoints; $^{*}$ denotes FD$_{\text{16k}}$ results computed by us.
S1, S2, and S3 denote the three training stages of Unite-Audio.
Best results are bold and underlined, and second-best results are bold.}
  \label{tab_01}
  \vspace{0.5em}

  \fontsize{8.5pt}{9.5pt}\selectfont
  \renewcommand{\arraystretch}{1.0}
  \begin{tabular}{l c c c c c c c c c c}

    \toprule
    Model & Params. & Text Cond. & Vocoder & NFE
    & FD$_{\text{OpenL3}}$ $\downarrow$
    & FD$_{\text{16k}}^{*}$ $\downarrow$
    & KL$_{\text{PaSST}}$ $\downarrow$
    & CLAP $\uparrow$
    & IS $\uparrow$
    & RTF $\downarrow$ \\
    \midrule

    AudioLDM2-Large $^{\ddagger}$ \cite{liu2023audioldm2}
      & 712M & T5 + CLAP & $\checkmark$ & 200
      & 108.3 & 40.03 & 1.81 & 0.419 & 7.90 & 2.589 \\

    Stable Audio Open $^{\ddagger}$ \cite{evans2024stableaudioopen}
      & 1056M & T5 & $\times$ & 100
      & 89.2 & 74.01 & 2.58 & 0.291 & 9.90 & 0.413 \\

    Tango2 $^{\ddagger}$ \cite{majumder2024tango2}
      & 866M & T5 & $\checkmark$ & 200
      & 108.4 & 45.45 & 1.11 & 0.447 & 9.00 & 0.781 \\

    TangoFlux-Base $^{\ddagger}$ \cite{hung2024tangoflux}
      & 515M & T5 & $\times$ & 50
      & 80.2 & 49.76 & 1.22 & 0.431 & \textbf{11.70} & 0.195 \\

    TangoFlux-RL $^{\ddagger}$ \cite{hung2024tangoflux}
      & 515M & T5 & $\times$ & 50
      & 75.1 & 52.28 & 1.15 & 0.480 & \underline{\textbf{12.20}} & 0.195 \\

    MeanAudio $^{\dagger}$ \cite{li2025meanaudio}
      & 120M & T5 + CLAP & $\checkmark$ & 25
      & 121.79 & 88.11 & 1.214 & 0.455 & 10.58 &  0.011 \\

    EzAudio $^{\dagger}$ \cite{hai2024ezaudio}
      & 875M & T5 + CLAP & $\times$ & 50
      & \underline{\textbf{38.26}} & 41.89 & 1.179 & 0.496 & 9.77 & 0.305  \\

    GenAU $^{\dagger}$ \cite{haji2026taming}
      & 1250M & T5 + CLAP & $\checkmark$ & 200
      & \textbf{68.92} & \underline{\textbf{31.94}} & 1.379 & 0.473 & 10.38 & 0.883 \\

    \midrule
    % \multicolumn{11}{l}{\textit{Joint training (ours)}} \\

    Unite-Audio \textbf{S1} (CFG=5.0)
      & \textbf{117M} & T5 & $\times$ & 32
      & 76.04 & 38.61 & 1.598 & 0.457 & 10.24 & 0.062 \\

    \addlinespace[1pt]

    Unite-Audio \textbf{S2} (CFG=5.0)
      & \textbf{117M} & T5 & $\times$ & 32
      & 73.46 & \textbf{33.41} & 1.205 & 0.502 & 10.41 & 0.062 \\

    \addlinespace[1pt]

    \rowcolor{stagefinal}
    Unite-Audio \textbf{S3} (CFG=5.0)
      & \textbf{117M} & T5 & $\times$ & 32
      & 84.76 & 43.25 & \textbf{1.107} & \underline{\textbf{0.534}} & 11.34 & 0.062 \\

    \rowcolor{stagefinal}
    Unite-Audio \textbf{S3} (CFG=2.0)
      & \textbf{117M} & T5 & $\times$ & 16
      & 83.22 & 39.38 & \underline{\textbf{1.057}} & \textbf{0.532} & 11.04 & 0.031 \\

    \rowcolor{stagefinal}
    Unite-Audio \textbf{S3} (CFG=2.0)
      & \textbf{117M} & T5 & $\times$ & 8
      & 84.88 & 42.41 & 1.079 & 0.530 & 10.85 & 0.016 \\

    \rowcolor{stagefinal}
    Unite-Audio \textbf{S3} (CFG=2.0)
      & \textbf{117M} & T5 & $\times$ & 4
      & 88.91 & 48.86 & 1.175 & 0.512 & 9.73 & 0.009 \\

    \bottomrule
  \end{tabular}
  \vspace{-1em}
\end{table*}

\vspace{-0.5em}
\subsection{Implementation Details}
\vspace{-0.5em}

For Stage I, we train for 240k iterations with a batch size of 24 using AdamW with a learning rate of $1\times10^{-4}$.
The EMA decay is set to 0.999. The context ratio $r_{\mathrm{ctx}}$ is sampled
from 0 to 30\%, and the reconstruction sampling probability $p_{\mathrm{rec}}$ is set to 0.5.
The reconstruction loss weights are
$\lambda_{\mathrm{G}}=1.0$,
$\lambda_{\mathrm{M}}=1.0$,
$\lambda_{\mathrm{GAN}}=0.1$, and
$\lambda_{\mathrm{F}}=5.0$. Adversarial training starts after 10k steps. Text conditioning is dropped with a probability of 10\% for classifier-free guidance (CFG).

For Stage II, we use AdamW with a learning rate of $5\times10^{-5}$, weight
decay of 0.01, and batch size 64. For Stage III, we use a learning rate of
$1\times10^{-6}$ and group size 4. Reward weights are
$\lambda_{\mathrm{CLAP}}=2/3$ and $\lambda_{\mathrm{PaSST}}=1/3$, with
$\epsilon=10^{-4}$ and $\beta=0.01$. For Unite-Audio, the CFG scale is 5.0
for 32-step sampling and 2.0 for lower-NFE settings. All experiments use two
NVIDIA H800 GPUs.

\vspace{-0.5em}
\subsection{Evaluation}
\vspace{-0.5em}

We follow the TangoFlux evaluation protocol \cite{hung2024tangoflux} on the
886-sample AudioCaps test split. We report OpenL3-based Fr\'echet Distance
(FD), PaSST-based KL divergence (KL), CLAP similarity, and Inception Score
(IS). FD measures the distributional similarity between generated and reference
audio, KL compares their PaSST-predicted event distributions, CLAP measures
text--audio semantic alignment, and IS reflects sample diversity and
specificity \cite{cramer2019openl3,koutini2021passt,salimans2016improved}.
We additionally report FD$_{\text{16k}}$, where both generated and reference
audio are resampled to 16 kHz before evaluation. Other metrics follow the same
evaluation models and settings as TangoFlux. RTF is averaged over ten runs of
10-second generation on a single NVIDIA H800 GPU.

\vspace{-0.5em}
\section{Results and Ablation Study}
\vspace{-0.5em}
\subsection{Performance}
\vspace{-0.5em}

Table~\ref{tab_01} compares Unite-Audio with existing TTA systems on the
AudioCaps-886 benchmark. With a 117M-parameter latent flow model, Unite-Audio achieves competitive overall
performance. Stage~II already shows strong generation before reward-based
post-training, while Flow-GRPO further improves semantic alignment and event
consistency. The final model achieves the highest CLAP score of 0.534 and the
lowest KL$_{\mathrm{PaSST}}$ of 1.057 among the compared systems, together with
a competitive IS. Although FD degrades after Flow-GRPO, Stage~II provides a
better trade-off in distributional fidelity. Unite-Audio also remains robust
under fewer sampling steps, with only modest degradation at 8 NFEs.

\vspace{-0.5em}
\subsection{Necessity of Noisy Partial-Context Flow Matching}
\vspace{-0.5em}

We compare three Stage-I checkpoints at 200k iterations, evaluated using EMA
weights on the same 886-sample AudioCaps test set, to examine how different
context-learning strategies affect generation and the learned encoder
representations. \emph{No context} removes the online representation from flow
matching, such that $\mathcal{L}_{\mathrm{FM}}$ has no direct gradient path to
the encoder. \emph{Clean context} restores this gradient path but introduces
clean latent information into otherwise noisy flow states, creating a mismatch
between the context and the current generative state. Our \emph{noisy context}
instead places the online representation at the same flow time as the generative
state.

Beyond generation metrics, we report effective rank to quantify latent
dimensional utilization, defined as
$r_{\mathrm{eff}}=\exp(-\sum_i p_i\log p_i)$, where
$p_i=\sigma_i^2/\sum_j\sigma_j^2$ denotes the normalized singular-value energy.
Semantic separability is further evaluated on ESC-50
\cite{piczak2015esc} using a linear probe, with classification accuracy
reported as Acc. Following Liu et al.~\cite{liu2026neural}, encoder-decoder
reconstruction quality is evaluated using multi-scale Mel distance and scale-invariant signal-to-distortion ratio (SI-SDR).
As shown in Table~\ref{tab:context_ablation}, noisy context performs best on
most generation and representation metrics, achieves the lowest Mel distance,
and maintains comparable SI-SDR, indicating improved representation learning
without compromising reconstruction quality.

\begin{table}[t]
\centering
\caption{Evaluation of latent representations learned with different
context-learning strategies at 200k iterations.}
\label{tab:context_ablation}
\vspace{0.3em}

\fontsize{8.5pt}{9.5pt}\selectfont
\setlength{\tabcolsep}{1.5pt}
\renewcommand{\arraystretch}{1.0}

\begin{tabular}{@{}lcccccccc@{}}
\toprule
Method
& FD$\downarrow$
& KL$\downarrow$
& IS$\uparrow$
& CLAP$\uparrow$
& Rank$\uparrow$
& Acc.$\uparrow$
& Mel$\downarrow$
& SI-SDR$\uparrow$ \\
\midrule

No context
& \textbf{79.87}
& 1.811
& 9.68
& 0.444
& 8.44
& 0.36
& 0.96
& 4.66 \\

Clean context
& 82.38
& 1.871
& 8.60
& 0.434
& 8.06
& 0.37
& 0.95
& \textbf{4.85} \\

\textbf{Noisy (ours)}
& 80.19
& \textbf{1.669}
& \textbf{10.37}
& \textbf{0.454}
& \textbf{8.90}
& \textbf{0.41}
& \textbf{0.94}
& 4.72 \\

\bottomrule
\end{tabular}

\vspace{-1.5em}
\end{table}

\vspace{-0.5em}
\section{Conclusion}
\vspace{-0.5em}

We introduced \textbf{Unite-Audio}, a framework that jointly learns continuous
audio tokenization and latent flow matching for TTA generation. Noise-matched
partial-context flow matching allows the generative objective to directly shape
the encoder while avoiding clean-context information unavailable at inference,
and full-generation training further aligns learning with generation from noise.
Flow-GRPO post-training further improves text-conditioned generation.
Experiments show that Unite-Audio achieves strong semantic fidelity with a compact 117M-parameter latent flow model, while retaining
competitive performance under few-step sampling.

\newpage

\bibliographystyle{IEEEbib}
\bibliography{references}

\end{document}